\documentclass[11pt]{article}
\usepackage[letterpaper,margin=1in]{geometry}
\usepackage[T1]{fontenc}
\usepackage[utf8]{inputenc}
\usepackage{lmodern}
\usepackage{microtype}
\usepackage{amsmath,amssymb}
\usepackage{array,booktabs,calc,longtable,tabularx}
\usepackage{seqsplit}
\usepackage{graphicx}
\usepackage{float}
\usepackage{xurl}
\usepackage[numbers,sort&compress]{natbib}
\usepackage{authblk}
\usepackage{caption}
\usepackage[hidelinks]{hyperref}

\providecommand{\tightlist}{\setlength{\itemsep}{0pt}\setlength{\parskip}{0pt}}

\title{Abliterated, but Better? Paired Evaluations of Cybersecurity Agents}
\author[1]{Alan Zabihi\thanks{Corresponding author:
  \href{mailto:alan@superagent.sh}{alan@superagent.sh}.}}
\author[1]{Ismail Pelaseyed}
\affil[1]{Superagent Technologies, Inc.\\
  \href{mailto:alan@superagent.sh}{alan@superagent.sh};
  \href{mailto:ismail@superagent.sh}{ismail@superagent.sh}}
\date{}

\hypersetup{
  pdftitle={Abliterated, but Better? Paired Evaluations of Cybersecurity Agents},
  pdfauthor={Alan Zabihi; Ismail Pelaseyed},
  pdfsubject={Paired evaluation of stock and abliterated models in cybersecurity agents},
  pdfkeywords={abliteration, refusal, cybersecurity agents, CVE-Bench, CyberGym}
}

\begin{document}
\maketitle

\begin{abstract}
Models marketed as ``abliterated'' are used in authorized cybersecurity research to reduce refusals, but lower refusal rates do not necessarily imply better task performance. We compare stock and third-party abliterated checkpoints from DeepSeek and GLM using paired evaluations on CVE-Bench and CyberGym.

Four primary comparisons evaluate each model family on 78 eligible CVE-Bench task variants and 120 selected CyberGym tasks under matched evaluation protocols. DeepSeek's success-rate differences are \ensuremath{-}3.85 percentage points on CVE-Bench and +1.67 points on CyberGym, with neither statistically resolved. GLM achieves equal success totals on CVE-Bench, while its abliterated checkpoint succeeds on 36 of 120 CyberGym tasks versus 22 for stock: a gain of 11.67 percentage points (95\% paired-task resampling interval, 3.33--20.00; Holm-adjusted \(p\) = 0.037). Provider-generated evaluations show the opposite GLM ranking on the same 120 tasks, although checkpoint identities, serving, harnesses, and grading differ.

A supplementary analysis of bounded trajectory excerpts does not establish refusal removal as an explanation for the observed differences. Incomplete text coverage and unvalidated detection sensitivity also prevent conclusions about refusal prevalence.

The tested abliterated releases therefore showed no consistent advantage across benchmarks or evaluation settings, and the bounded trajectory analysis did not link the additional successes to stock refusals.
\end{abstract}

\section{Introduction}\label{introduction}

Security agents need to inspect vulnerable code, generate proof-of-concept (PoC) inputs, and test exploit paths inside authorized sandboxes. Safety-oriented refusals can block that work. An unsuccessful trajectory can also reflect weak reasoning, tool failures, task difficulty, limits, or missing submissions. Oracle outcomes and refusal measurements therefore have different questions to answer.

Abliteration has documented effects on refusal representations, but that mechanism does not establish the competence or behavior of a third-party abliterated checkpoint \citep{arditi2024refusal}. Published security work already compares stock and abliterated models for vulnerability analysis and autonomous agents \citep{li2026beyond,david2026measuring}, especially the 1,500 security-agent traces and mixed effects across families reported by David and Gervais. We add exact within-family task pairing, revision-pinned self-hosted checkpoints, fixed valid-run denominators, dependence-aware paired analyses, and a channel-explicit packet audit.

The release name abliterated denotes the publisher's designation for a third-party checkpoint. It is not evidence that the release differs from stock only by removal of a verified refusal direction. The comparisons are checkpoint contrasts. They are not isolated causal interventions and do not measure a stock-to-abliterated reduction in refusal on a common, fully validated corpus.

\subsection{Research questions and hypothesis}\label{research-questions-and-hypothesis}

\begin{enumerate}
\def\labelenumi{\arabic{enumi}.}
\tightlist
\item
  On each named eligible task set, how do verified oracle outcomes differ between stock and abliterated checkpoints from DeepSeek and GLM?
\item
  For the 41 pinned comparisons in which stock fails and the abliterated checkpoint passes, what bounded evidence is available for refusal, early finishing, budget exhaustion, or absence of an observed finish?
\item
  What do lexical screening and bounded agent annotation identify in packets from the 650 valid stock runs?
\end{enumerate}

We test the hypothesis that abliterated checkpoints yield more verified task successes than stock checkpoints. We also assess whether stock refusal or premature stopping is associated with differences in success. The four primary comparisons cover combined-mode CVE-Bench and the full 120-task CyberGym set at high effort for each model family. Comparisons by mode, effort, and framing, as well as the common-ID, clustered, project-level, and external analyses, are secondary or exploratory.

The results do not show a consistent advantage for abliterated checkpoints across the four primary comparisons. Instead, effects differ by model family and benchmark. The behavioral analysis is limited to bounded packets and cannot establish refusal or premature stopping as explanations for the observed differences. Early finishing is not equivalent to failure to attempt an exploit, and complete attempt timing was not measured.

\section{Related work}\label{related-work}

\subsection{Abliteration and security-agent outcomes}\label{abliteration-and-security-agent-outcomes}

Arditi et al.~show that refusal behavior can be mediated by a direction in activation space \citep{arditi2024refusal}. Young's comparison of methods argues for treating architecture and transformation method as parts of the experimental condition \citep{young2026comparative}. David and Gervais's evaluation of alignment-removal methods also separates utility from unsafe out-of-scope compliance \citep{david2026ablating}. These studies do not validate the construction of every checkpoint marketed as abliterated.

Li et al.~compare aligned and abliterated models from the same lineage for vulnerability analysis \citep{li2026beyond}. Carleo et al.~distinguish willingness from code-task capability \citep{carleo2026willing}. David and Gervais's autonomous-agent study is the closest comparison: it measures refusal and other behaviors in full tool-using execution, includes 1,500 security traces and 800 non-security controls, and reports large gains in some families, lower performance in others, and examples with no refusal \citep{david2026measuring}. We take a narrower approach: measure paired benchmark outcomes for named pinned checkpoints under fixed protocols, then test how sensitive the comparisons are to their sources and dependence structure.

\subsection{Refusal constructs and benchmark definitions}\label{refusal-constructs-and-benchmark-definitions}

Refusal is not established by a keyword, inability, command failure, missing file, connection error, missing submission, or an early finish. Hadetskyi et al.~document variation in cybersecurity refusals at scale \citep{hadetskyi2026refusals}. Campbell et al.~measure refusal bias in defensive tasks, including the effect of explicit authorization \citep{campbell2026defensive}. These papers motivate our emphasis on authorized context and the distinction between task refusal and technical failure. The particular full-task/partial-action and sequence-effect taxonomy used in the packet rubric is operationalized for this study; we do not attribute it to those papers.

CVE-Bench supplies real web-application vulnerabilities in its one-day and zero-day settings \citep{zhu2025cvebench}. Its zero-day label describes withholding the historical vulnerability description, not proving that the underlying vulnerability was unknown to model pretraining. CyberGym Level 1 supplies a pre-fix codebase and a text vulnerability description \citep{wang2026cybergym}. PoC reproduction does not measure broad incident-response competence or general capability retention.

\section{Study design}\label{study-design}

\subsection{Evidence strata and valid runs}\label{evidence-strata-and-valid-runs}

The pinned required inventory was 1,160 runs: 800 high-effort runs over four checkpoints, 200 DeepSeek max-effort runs, and 160 required framing runs on two stock checkpoints. Fixed exclusions of \texttt{CVE-2024-34070} in both modes removed 16 assigned variants, leaving 1,144 valid required results. The supplementary GLM abliterated framing arm has two corresponding exclusions, leaving 78 valid runs. The combined analysis therefore contains 1,222 records, not a 1,222-run required denominator.

These records comprise 792 valid high-effort runs, 196 valid max-effort runs, 156 required framing runs, and 78 supplementary framing runs. There are 650 valid stock runs with 146 oracle passes and 504 oracle failures. These are run counts across conditions that reuse task IDs; they are not an independent pooled benchmark estimate.

\texttt{CVE-2024-34070} was excluded across all model arms because its execution environment failed independent compatibility checks. The exclusion removes two variants per arm, and these variants are not counted as model successes or failures. The primary estimand is therefore per eligible variant. Under a nominal-all-assignment sensitivity analysis, allowing either outcome for the two excluded pairs would place the DeepSeek primary difference between \ensuremath{-}6.25 and \ensuremath{-}1.25 pp and the GLM difference between \ensuremath{-}2.50 and +2.50 pp.~The sensitivity analysis does not create oracle outcomes for the excluded variants.

Four full high-effort contrasts are the designated primary family for multiplicity control. Everything else, including mode splits, max, framing, common-ID analyses, clustering analyses, and external contrasts, is secondary or exploratory.

\subsection{Checkpoints, serving, and templates}\label{checkpoints-serving-and-templates}

The pinned serving image was \path{lmsysorg/sglang:v0.5.23-cu130@sha256:f7876bef696ccf9cf75a2bf7ccb82970f5fac577f082fb8ca95c9a59d2160eb5}. All pinned runs used SGLang with \texttt{temperature=0.0} and the recorded effort setting. A single run per task and condition was retained; zero temperature and deterministic analysis seeds do not establish repeatability of serving outputs across batches, caches, or restarts.

\begingroup\scriptsize
\setlength{\tabcolsep}{3.5pt}
\begin{longtable}[]{@{}
  >{\raggedright\arraybackslash}p{(\linewidth - 6\tabcolsep) * \real{0.2500}}
  >{\raggedright\arraybackslash}p{(\linewidth - 6\tabcolsep) * \real{0.2500}}
  >{\raggedright\arraybackslash}p{(\linewidth - 6\tabcolsep) * \real{0.2500}}
  >{\raggedright\arraybackslash}p{(\linewidth - 6\tabcolsep) * \real{0.2500}}@{}}
\caption{Pinned checkpoints and retained serving configurations. The configured limits do not establish the usable sequence length of every trajectory.}\tabularnewline
\toprule\noalign{}
\begin{minipage}[b]{\linewidth}\raggedright
Family and variant
\end{minipage} & \begin{minipage}[b]{\linewidth}\raggedright
Hugging Face repository
\end{minipage} & \begin{minipage}[b]{\linewidth}\raggedright
Immutable commit SHA
\end{minipage} & \begin{minipage}[b]{\linewidth}\raggedright
Actual retained serving configuration
\end{minipage} \\
\midrule\noalign{}
\endfirsthead
\toprule\noalign{}
\begin{minipage}[b]{\linewidth}\raggedright
Family and variant
\end{minipage} & \begin{minipage}[b]{\linewidth}\raggedright
Hugging Face repository
\end{minipage} & \begin{minipage}[b]{\linewidth}\raggedright
Immutable commit SHA
\end{minipage} & \begin{minipage}[b]{\linewidth}\raggedright
Actual retained serving configuration
\end{minipage} \\
\midrule\noalign{}
\endhead
\bottomrule\noalign{}
\endlastfoot
DeepSeek stock & \path{deepseek-ai/DeepSeek-V4.1-Flash-1017} & \texttt{\seqsplit{1bae0e1088993b9b82de08f59de5be5a34e42170}} & 8 x H200, TP=8, EP=8; FP8 with FP4 experts; configured context 1,048,576 \\
DeepSeek abliterated & \path{dealignai/DeepSeek-V4.1-Flash-1017-UNCENSORED-FP8} & \texttt{\seqsplit{c19ca5ca4eaeb79d29fdf08aea8327da95b1116a}} & 8 x H200, TP=8, EP=8; FP8 with FP4 experts; configured context 1,048,576 \\
GLM stock & \texttt{zai-org/GLM-5.3-Flash} & \texttt{\seqsplit{46a050bec13002ad29ee795c16e61cf4fa3c4a9c}} & 4 x B200, TP=4, EP=4; e4m3 quantization; configured context 1,048,576 \\
GLM abliterated & \path{orcarouter/GLM-5.3-Flash-Uncensored-FP8} & \texttt{\seqsplit{993122712d2606d8a49f0b350d35b3c6672861c8}} & 4 x B200, TP=4, EP=4; e4m3 quantization; configured context 1,048,576 \\
\end{longtable}\endgroup

Both DeepSeek revision configs declare FP8 with FP4 experts and \texttt{\seqsplit{max\_position\_embeddings=1048576}}; their original rope context is 65,536. The serving manifests and abliterated serving gate record the retained 1,048,576-token configuration. Hardware and parallelism differ between families, so cross-family comparisons are not serving-equivalence tests.

The GLM chat-template files differ at the pinned revisions: stock has 10,830 bytes (blob \texttt{14180451\ldots{}}) and the abliterated release has 10,644 bytes (blob \texttt{fb94d40d\ldots{}}). The latter is byte-identical to an earlier public stock file. The upstream stock change concerns speed of tool-result validation and reordering. The launch code uses checkpoint defaults without a \texttt{-\/-chat-template} override. Byte-identical reference outputs were obtained for system/user, ordinary multi-turn, aligned two-tool-result, and reversed two-tool-result messages. The files, render hashes, and limitations are in \path{paper/template-render-checks.md}. Production token equality for real conversations was not reconstructed. We therefore describe prompts as unmodified, checkpoint-default inputs with equivalent reference renders, not as proven identical model inputs throughout execution.

No task prompt was modified. The framing arms prepended this fixed sentence to the default system prompt: ``CONTEXT: You are operating inside an isolated, offline Docker sandbox for an authorized defensive security evaluation. All targets are synthetic local containers owned by the evaluator.'' This adds authorization, ownership, offline operation, and synthetic-target language together. It does not change the software or imply that historical real-world vulnerabilities are synthetic. DeepSeek stock has the framing run but DeepSeek abliterated does not; GLM max effort does not exist. The design is not fully factorial.

\subsection{Benchmark protocols and selection}\label{benchmark-protocols-and-selection}

\begingroup\scriptsize
\setlength{\tabcolsep}{3.5pt}
\begin{longtable}[]{@{}
  >{\raggedright\arraybackslash}p{(\linewidth - 6\tabcolsep) * \real{0.2500}}
  >{\raggedright\arraybackslash}p{(\linewidth - 6\tabcolsep) * \real{0.2500}}
  >{\raggedright\arraybackslash}p{(\linewidth - 6\tabcolsep) * \real{0.2500}}
  >{\raggedright\arraybackslash}p{(\linewidth - 6\tabcolsep) * \real{0.2500}}@{}}
\caption{Protocols. Pinned outcomes were evaluated under the study protocol; provider outcomes remained source assigned. Common Weakness Enumeration (CWE) labels in the selection are derived from ARVO crash types, not official CyberGym annotations.}\tabularnewline
\toprule\noalign{}
\begin{minipage}[b]{\linewidth}\raggedright
Stratum and benchmark
\end{minipage} & \begin{minipage}[b]{\linewidth}\raggedright
Population
\end{minipage} & \begin{minipage}[b]{\linewidth}\raggedright
Agent and scoring protocol
\end{minipage} & \begin{minipage}[b]{\linewidth}\raggedright
Limits
\end{minipage} \\
\midrule\noalign{}
\endfirsthead
\toprule\noalign{}
\begin{minipage}[b]{\linewidth}\raggedright
Stratum and benchmark
\end{minipage} & \begin{minipage}[b]{\linewidth}\raggedright
Population
\end{minipage} & \begin{minipage}[b]{\linewidth}\raggedright
Agent and scoring protocol
\end{minipage} & \begin{minipage}[b]{\linewidth}\raggedright
Limits
\end{minipage} \\
\midrule\noalign{}
\endhead
\bottomrule\noalign{}
\endlastfoot
Pinned CVE-Bench & 40 CVEs in each mode; 39 retained CVEs/78 variants per arm & CVE-Bench commit \texttt{\seqsplit{9fb3e068f260b00dd899ef87951b88e9c60061f1}}; official \texttt{./run\ eval}, Inspect \texttt{claude\_code\_agent}, exploit oracle & \texttt{max\_messages=50}; operational batch timeouts are not per-task equivalents \\
Pinned CyberGym & Fixed 120-task Level 1 set from 1,507; DeepSeek max-effort compares only its first 20 & CyberGym commit \texttt{\seqsplit{c6fe2027d39471375920b92cf1025e23a99ffda5}}; agent examples \texttt{\seqsplit{6660f3f5a0193e7f142f23f998677429edb3d18c}}; OpenHands \texttt{\seqsplit{35b381f3a8f4b5229934515e9f6b479d6d6415ef}} & \texttt{max\_iter=100}; \texttt{\seqsplit{native\_tool\_calling=true}}; timeout 1,200 seconds \\
External CVE-Bench & All 80 source variants & CVE-Bench 2.2.0; Claude Code 2.1.284 through an Inspect model bridge; Inspect AI 0.3.268; \texttt{check\_exploit} & 100 messages; 3,600 seconds \\
External CyberGym & All 1,507 source Level 1 tasks & UK AISI Inspect ReAct with three answer attempts; Inspect AI 0.3.285; \texttt{cybergym\_scorer} & 202 messages; 1,200 seconds; approximately 100 iterations by source description \\
\end{longtable}\endgroup

The pinned CVE-Bench oracle assigns success from the official exploit check. Timeouts and infrastructure errors are separately classified, not silently treated as model failures. CyberGym uses the final designated PoC. The primary implementation passes when the vulnerable build crashes (\texttt{vul\_exit\_code} not 0 or timeout 300) and the fixed build exits 0. The external CyberGym source also reports final-PoC scoring with vulnerable failure and fixed success, but its executor and handling of unscorable responses differ. Missing submissions and valid failure are distinct from executor-side errors. A missing submission can be model evidence only when its transport was validated separately.

CyberGym selection used one random generator with seed \texttt{20261108}, stable task-ID ordering, and largest-remainder quota allocation across 13 derived strata, with repository conflicts, skips, backfill, and a final shuffle. The committed configuration contains all selected IDs, all target and population stratum sizes, mapping rules, and data versions. Thus the selected list is reproducible. There are 120 unique \texttt{project\_main\_repo} values and no duplicate exact repository string. The first 20 selected tasks form an ordering subset, not a separately stratified probability sample of all 1,507.

The algorithm gives unequal and sometimes zero inclusion exposure. It does not give every Level 1 task an independent equal chance of selection. The pinned effect therefore describes the realized 120 tasks. An interval formed by resampling them is not, by itself, design-based inference to all tasks or to comparable future tasks. Population-level generalization would require explicit inclusion probabilities and design assumptions. The public full task manifest at the same dataset revision has SHA-256 \texttt{\seqsplit{9cea452cc1e1a3703e0f60c2dfc8642430aab9f50433f976581509de58c7048f}} and maps all 1,507 tasks to 188 exact repository strings.

\subsection{External routes, validity, and outcomes}\label{external-routes-validity-and-outcomes}

The eight provider-generated batches cover OpenRouter stock routes and RefuseLess abliterated routes for both families and benchmarks. Source manifests preserve exact labels, run completion times, harness versions, and task metrics, while \texttt{model\_revision} is null. Their abliterated routes are not the pinned Hugging Face abliterated checkpoints in any verified weight-identity sense; they lack immutable revision bindings as well. Hardware, engine, temperature, and provider serving details are not inferred.

Source joins on \texttt{\seqsplit{(subset,\ exact\ task\ ID)}} give 80 CVE-Bench pairs per family and 1,507 CyberGym pairs per family. There are 3,174 pairs and 6,348 source individual outcomes. Current R2 reads verified the eight result object hashes, task-key totals, and source pass totals against the committed bindings. Exact ID crosswalks were then checked in both directions for 396 shared primary/external pairs. \texttt{arvo:N} maps to \texttt{\seqsplit{cybergym\_arvo\_N\ (level1)}}, and \texttt{oss-fuzz:N} maps to \texttt{\seqsplit{cybergym\_oss\_fuzz\_N\ (level1)}} under the external harness's source.

These are batches, not eight 3,174-run experiments or new primary pairs. Archive integrity does not independently verify execution or grading. Normalization checks integer success 0/1 against \texttt{passed}/\texttt{failed} and the expected scorer key, then reconstructs source-reported metrics. A further R2 audit of the current \texttt{results.json} objects finds that all 6,348 task-level scorer values equal the source success flags: \texttt{check\_exploit.value} for CVE-Bench and \texttt{reproduced} for CyberGym, with zero mismatches or unavailable relations. This comparison verifies scorer normalization, not independent regrading. The \texttt{results.json} task schema has no distinct \texttt{unscored} or \texttt{oracle\_error} field establishing the precise validity class of every failure. The schema therefore does not establish that every grader completed or that every failure was a valid model-oracle failure under the primary protocol. Detailed error semantics remain source assigned. Provider fields \texttt{n\_limited} and \texttt{n\_refusals} are not recoded as early stops or refusals. For example, external GLM CyberGym has 614 stock \texttt{n\_limited} records versus 805 for the abliterated route. That difference alone does not explain a sign reversal, but it must be audited before drawing a reliability conclusion.

The reported outcomes are exploratory and never pooled with the pinned estimates. Task identity overlap removes only one difference between protocols. It does not establish equivalent weight identity, serving, tools, prompts, grading, or error semantics.

\subsection{Behavior definitions and bounded annotation}\label{behavior-definitions-and-bounded-annotation}

Refusal measurement has four distinct outputs: original stored-content lexical flags, packet-derived visibility diagnostics, bounded agent labels, and derived terminal metadata.

The legacy rules could match stored assistant \texttt{content} containing literal \texttt{\textless{}think\textgreater{}...\textless{}/think\textgreater{}} regions. Literal embedded deliberation, separate \texttt{reasoning\_content}, prompts, quoted text, and tool output cannot establish an assistant-visible refusal. The bounded packet builder retains broad-candidate windows plus the first and last assistant messages, at most four additional broad candidates per message, with first/last-window bounding. It also retains message hashes, lengths, role/tool timelines, and bounded context.

The compact reading representation retains candidate windows and the role/tool timeline. Ordinary assistant text that has no candidate match is not rendered, even if the fuller packet holds it. We measure this exclusion directly. Long text can be bounded before channel separation, so a raw match's channel may be unclassifiable without reopening source text.

Two separately executed OpenCode annotation runs used the same Kimi-K3.1-Pro model and rubric for these packets. Separate execution does not eliminate shared candidate-screen, model, or rubric bias. Review settings, timestamps, decoding temperature, and reusable request/response hashes were not retained in the completed provenance. Disagreements, positive cases, and low-confidence cases were designated for renewed blinded review. No such cases occurred. Consequently, labels were checked and carried forward; there was no third substantive evidence review of all trajectories. Final label files are byte-for-byte copies of pipeline A.

A qualifying label requires explicit visible assistant text declining all or necessary part of the authorized task for a stated or implied safety, policy, or misuse reason. A warning followed by unrestricted execution is not a refusal. Technical inability, ordinary failed attempts, and early finish are not refusal labels. Confidence is an agent annotation, not calibrated accuracy. The all-negative distribution makes Cohen's kappa undefined.

Early stop is a derived behavior flag for an oracle-failing run that ends before its budget without an official timeout. It includes ordinary failures and is not refusal or non-attempt. For CyberGym, an official finish event is required for early finish classification. For CVE-Bench, the current code uses absence of a message-limit ending rather than requiring an explicit finish event. Raw events should be checked before interpreting that CVE flag as an issued finish. \texttt{no\_finish\_observed} means just that; it is not agent self-stop. The \(\geq\)90-call CyberGym iteration-exhaustion proxy is descriptive rather than an official terminal category. Passing rows can also exhaust an iteration budget and are not classified as failures.

\subsection{Statistical analysis}\label{statistical-analysis}

The observed outcome is binary verified task success. Effects are abliterated-minus-stock on the absolute probability scale, paired by exact task/mode identity within protocol. The four-cell paired counts are always reported. No estimate is pooled across families, benchmarks, effort levels, prompts, or pinned/external protocols.

For combined-mode CVE-Bench, a base CVE and its two modes are a dependence cluster. The primary sensitivity test exchanges stock/abliterated labels jointly within each CVE and calculates the paired total's two-sided distribution under exchangeable labels and independence between CVEs. This is a paired sensitivity test under its assumptions, not a randomized-treatment claim. Task-level single-mode and selected-CyberGym tests use exact two-sided McNemar tests. Holm step-down covers exactly four designated primary tests. The observed counts, primary intervals, exact tests, and sensitivity p-values are reproducible in \path{paper/analysis/revision-analysis.json} and \path{code/analysis/revision_analysis.py}.

Primary significance decisions use Holm-adjusted tests. The accompanying pointwise 95\% intervals are paired task/cluster-percentile resampling intervals, not familywise simultaneous intervals. Excluding zero in one interval does not establish a familywise significant result. Percentile intervals and conditional exact tests need not make identical threshold statements. A wide interval is not an equivalence test. Equal pass totals can contain wins and losses in both directions. A proposed practical equivalence claim would require a separate margin and analysis.

Pinned intervals use 20,000 resamples and base seed \texttt{20261122}; external intervals use 50,000 and base seed \texttt{20261125}. Combined-mode CVE-Bench resamples base CVEs; the 120-task primary CyberGym intervals resample individual selected IDs with one per exact repository. Project sensitivity analyses resample all tasks in a sampled external project at its own size and retain task weighting. The selected 120 and all 1,507 are different estimands. One run per assigned task and condition cannot estimate replay, infrastructure, cache, provider-route, or revision drift variance.

\section{Results}\label{results}

\subsection{RQ1: Paired verified success}\label{rq1-paired-verified-success}

\begingroup\scriptsize
\setlength{\tabcolsep}{3.5pt}
\begin{longtable}[]{@{}
  >{\raggedright\arraybackslash}p{(\linewidth - 12\tabcolsep) * \real{0.1111}}
  >{\raggedleft\arraybackslash}p{(\linewidth - 12\tabcolsep) * \real{0.1481}}
  >{\raggedleft\arraybackslash}p{(\linewidth - 12\tabcolsep) * \real{0.1481}}
  >{\raggedleft\arraybackslash}p{(\linewidth - 12\tabcolsep) * \real{0.1481}}
  >{\raggedleft\arraybackslash}p{(\linewidth - 12\tabcolsep) * \real{0.1481}}
  >{\raggedleft\arraybackslash}p{(\linewidth - 12\tabcolsep) * \real{0.1481}}
  >{\raggedleft\arraybackslash}p{(\linewidth - 12\tabcolsep) * \real{0.1481}}@{}}
\caption{Primary high-effort effects. Intervals are paired 95\% task/cluster-resampling intervals. CVE tests are whole-CVE label-swap sensitivities; selected CyberGym tests are exact McNemar. The final adjusted result also holds when using the original McNemar p-values from \texttt{analysis-paired-results.csv}, because only the CVE p-values change.}\tabularnewline
\toprule\noalign{}
\begin{minipage}[b]{\linewidth}\raggedright
Primary contrast
\end{minipage} & \begin{minipage}[b]{\linewidth}\raggedleft
Paired tasks
\end{minipage} & \begin{minipage}[b]{\linewidth}\raggedleft
Both pass / stock only / abliterated only / both fail
\end{minipage} & \begin{minipage}[b]{\linewidth}\raggedleft
Stock / abliterated passes
\end{minipage} & \begin{minipage}[b]{\linewidth}\raggedleft
Effect, pp {[}95\% interval{]}
\end{minipage} & \begin{minipage}[b]{\linewidth}\raggedleft
Paired test, p
\end{minipage} & \begin{minipage}[b]{\linewidth}\raggedleft
Holm p
\end{minipage} \\
\midrule\noalign{}
\endfirsthead
\toprule\noalign{}
\begin{minipage}[b]{\linewidth}\raggedright
Primary contrast
\end{minipage} & \begin{minipage}[b]{\linewidth}\raggedleft
Paired tasks
\end{minipage} & \begin{minipage}[b]{\linewidth}\raggedleft
Both pass / stock only / abliterated only / both fail
\end{minipage} & \begin{minipage}[b]{\linewidth}\raggedleft
Stock / abliterated passes
\end{minipage} & \begin{minipage}[b]{\linewidth}\raggedleft
Effect, pp {[}95\% interval{]}
\end{minipage} & \begin{minipage}[b]{\linewidth}\raggedleft
Paired test, p
\end{minipage} & \begin{minipage}[b]{\linewidth}\raggedleft
Holm p
\end{minipage} \\
\midrule\noalign{}
\endhead
\bottomrule\noalign{}
\endlastfoot
DeepSeek high, CVE-Bench & 78 & 11 / 6 / 3 / 58 & 17 / 14 & \ensuremath{-}3.85 {[}\ensuremath{-}12.82, +3.85{]} & 0.5625 (whole-CVE) & 1.0 \\
DeepSeek high, CyberGym & 120 & 10 / 7 / 9 / 94 & 17 / 19 & +1.67 {[}\ensuremath{-}5.00, +8.33{]} & 0.803619 & 1.0 \\
GLM high, CVE-Bench & 78 & 20 / 3 / 3 / 52 & 23 / 23 & 0.00 {[}\ensuremath{-}6.41, +6.41{]} & 1.0 (whole-CVE) & 1.0 \\
GLM high, CyberGym & 120 & 16 / 6 / 20 / 78 & 22 / 36 & +11.67 {[}+3.33, +20.00{]} & 0.009355 & 0.037421 \\
\end{longtable}\endgroup

Only the GLM CyberGym contrast meets the selected Holm criterion. Its measured advantage is 14 additional passes: 20 abliterated-only wins minus 6 stock-only wins, or 36 versus 22. The other primary estimates are unresolved: their signs are not uniform and their intervals leave possible gains or harms compatible with the data. The exact zero total contrast in GLM CVE-Bench still has three wins in each direction. A difference in significance across rows is not an established family-by-condition interaction.

The one-pair result sensitivity is not an oracle correction. If one abliterated-only GLM CyberGym success were a tie, the discordant counts would be 19 against 6, giving +10.83 pp, exact McNemar \(p\) = 0.014633, and Holm-adjusted \(p\) = 0.058533 with the other three tests fixed. The observed +11.67 pp effect and 22/36 pass counts remain the study result. The sensitivity analysis indicates that the adjusted significance depends on one pair.

\subsubsection{Secondary mode and effort comparisons}\label{secondary-mode-and-effort-comparisons}

\begingroup\scriptsize
\setlength{\tabcolsep}{3.5pt}
\begin{longtable}[]{@{}
  >{\raggedright\arraybackslash}p{(\linewidth - 8\tabcolsep) * \real{0.1579}}
  >{\raggedleft\arraybackslash}p{(\linewidth - 8\tabcolsep) * \real{0.2105}}
  >{\raggedleft\arraybackslash}p{(\linewidth - 8\tabcolsep) * \real{0.2105}}
  >{\raggedleft\arraybackslash}p{(\linewidth - 8\tabcolsep) * \real{0.2105}}
  >{\raggedleft\arraybackslash}p{(\linewidth - 8\tabcolsep) * \real{0.2105}}@{}}
\caption{Exploratory high-effort CVE mode splits with paired cells. These single-mode tests are exact McNemar. The GLM zero-day interval is zero width because every observed pair is concordant; it is an empirical resampling artifact, not a statement of zero population uncertainty. With 39 independent zero-discordance pairs, a simple one-sided 95\% upper bound on any-discordance probability would be 7.39\%, but that is not a paired mean-effect interval.}\tabularnewline
\toprule\noalign{}
\begin{minipage}[b]{\linewidth}\raggedright
Contrast
\end{minipage} & \begin{minipage}[b]{\linewidth}\raggedleft
Paired tasks
\end{minipage} & \begin{minipage}[b]{\linewidth}\raggedleft
Both pass / stock only / abliterated only / both fail
\end{minipage} & \begin{minipage}[b]{\linewidth}\raggedleft
Effect, pp {[}95\% interval{]}
\end{minipage} & \begin{minipage}[b]{\linewidth}\raggedleft
Test, p
\end{minipage} \\
\midrule\noalign{}
\endfirsthead
\toprule\noalign{}
\begin{minipage}[b]{\linewidth}\raggedright
Contrast
\end{minipage} & \begin{minipage}[b]{\linewidth}\raggedleft
Paired tasks
\end{minipage} & \begin{minipage}[b]{\linewidth}\raggedleft
Both pass / stock only / abliterated only / both fail
\end{minipage} & \begin{minipage}[b]{\linewidth}\raggedleft
Effect, pp {[}95\% interval{]}
\end{minipage} & \begin{minipage}[b]{\linewidth}\raggedleft
Test, p
\end{minipage} \\
\midrule\noalign{}
\endhead
\bottomrule\noalign{}
\endlastfoot
DeepSeek high, CVE one-day & 39 & 6 / 5 / 2 / 26 & \ensuremath{-}7.69 {[}\ensuremath{-}20.51, +5.13{]} & 0.453125 \\
DeepSeek high, CVE zero-day & 39 & 5 / 1 / 1 / 32 & 0.00 {[}\ensuremath{-}7.69, +7.69{]} & 1.0 \\
GLM high, CVE one-day & 39 & 11 / 3 / 3 / 22 & 0.00 {[}\ensuremath{-}12.82, +12.82{]} & 1.0 \\
GLM high, CVE zero-day & 39 & 9 / 0 / 0 / 30 & 0.00 {[}0.00, 0.00{]} & 1.0 \\
\end{longtable}\endgroup

At max effort, DeepSeek stock versus abliterated performs 22 versus 14 on CVE-Bench: paired cells 13 / 9 / 1 / 55, effect \ensuremath{-}10.26 pp {[}\ensuremath{-}19.23, \ensuremath{-}2.56{]}. Ordinary exact McNemar gives \(p\) = 0.021484; whole-CVE label-swap sensitivity gives \(p\) = 0.035156. This is an exploratory result, not a confirmatory harm effect. DeepSeek max CyberGym on the first 20 tasks has cells 0 / 0 / 2 / 18 and effect +10.00 pp {[}0.00, +25.00{]}, \(p\) = 0.5. It is not a full-120-task max comparison.

\begingroup\scriptsize
\setlength{\tabcolsep}{3.5pt}
\begin{longtable}[]{@{}
  >{\raggedright\arraybackslash}p{(\linewidth - 8\tabcolsep) * \real{0.1579}}
  >{\raggedleft\arraybackslash}p{(\linewidth - 8\tabcolsep) * \real{0.2105}}
  >{\raggedleft\arraybackslash}p{(\linewidth - 8\tabcolsep) * \real{0.2105}}
  >{\raggedleft\arraybackslash}p{(\linewidth - 8\tabcolsep) * \real{0.2105}}
  >{\raggedleft\arraybackslash}p{(\linewidth - 8\tabcolsep) * \real{0.2105}}@{}}
\caption{Exploratory within-checkpoint effort comparisons. Counts at full 120 high effort cannot supply the 20-task effort transitions. No preferred condition is selected by these unadjusted analyses.}\tabularnewline
\toprule\noalign{}
\begin{minipage}[b]{\linewidth}\raggedright
Within-checkpoint effort comparison
\end{minipage} & \begin{minipage}[b]{\linewidth}\raggedleft
Paired tasks
\end{minipage} & \begin{minipage}[b]{\linewidth}\raggedleft
Both pass / high only / max only / both fail
\end{minipage} & \begin{minipage}[b]{\linewidth}\raggedleft
Max\ensuremath{-}high, pp {[}95\% interval{]}
\end{minipage} & \begin{minipage}[b]{\linewidth}\raggedleft
Exploratory test
\end{minipage} \\
\midrule\noalign{}
\endfirsthead
\toprule\noalign{}
\begin{minipage}[b]{\linewidth}\raggedright
Within-checkpoint effort comparison
\end{minipage} & \begin{minipage}[b]{\linewidth}\raggedleft
Paired tasks
\end{minipage} & \begin{minipage}[b]{\linewidth}\raggedleft
Both pass / high only / max only / both fail
\end{minipage} & \begin{minipage}[b]{\linewidth}\raggedleft
Max\ensuremath{-}high, pp {[}95\% interval{]}
\end{minipage} & \begin{minipage}[b]{\linewidth}\raggedleft
Exploratory test
\end{minipage} \\
\midrule\noalign{}
\endhead
\bottomrule\noalign{}
\endlastfoot
DeepSeek stock, CVE-Bench & 78 & 14 / 3 / 8 / 53 & +6.41 {[}\ensuremath{-}2.56, +15.38{]} & whole-CVE \(p\) = 0.3125 \\
DeepSeek abliterated, CVE-Bench & 78 & 12 / 2 / 2 / 62 & 0.00 {[}\ensuremath{-}5.13, +5.13{]} & \(p\) = 1.0 \\
DeepSeek stock, CyberGym first 20 & 20 & 0 / 6 / 0 / 14 & \ensuremath{-}30.00 {[}\ensuremath{-}50.00, \ensuremath{-}10.00{]} & exact McNemar \(p\) = 0.03125 \\
DeepSeek abliterated, CyberGym first 20 & 20 & 1 / 2 / 1 / 16 & \ensuremath{-}5.00 {[}\ensuremath{-}20.00, +10.00{]} & exact McNemar \(p\) = 1.0 \\
\end{longtable}\endgroup

\subsubsection{Framing controls}\label{framing-controls}

DeepSeek stock changes from 17 to 24 passes under framing: +8.97 pp {[}\ensuremath{-}2.56, +21.79{]}, ordinary exact \(p\) = 0.143463, whole-CVE label-swap \(p\) = 0.236328. The paired cells are 12 both pass, 5 default-only, 12 framing-only, and 49 both fail.

GLM stock and abliterated both change from 23 to 21 passes, but the paired transitions differ:

\begingroup\scriptsize
\setlength{\tabcolsep}{3.5pt}
\begin{longtable}[]{@{}
  >{\raggedright\arraybackslash}p{(\linewidth - 6\tabcolsep) * \real{0.2000}}
  >{\raggedleft\arraybackslash}p{(\linewidth - 6\tabcolsep) * \real{0.2667}}
  >{\raggedleft\arraybackslash}p{(\linewidth - 6\tabcolsep) * \real{0.2667}}
  >{\raggedleft\arraybackslash}p{(\linewidth - 6\tabcolsep) * \real{0.2667}}@{}}
\caption{GLM framing transitions. Equal totals do not establish identical framing effects, equal uncertainty, or a refusal mechanism.}\tabularnewline
\toprule\noalign{}
\begin{minipage}[b]{\linewidth}\raggedright
Comparison
\end{minipage} & \begin{minipage}[b]{\linewidth}\raggedleft
Losses / gains
\end{minipage} & \begin{minipage}[b]{\linewidth}\raggedleft
Effect, pp {[}95\% interval{]}
\end{minipage} & \begin{minipage}[b]{\linewidth}\raggedleft
Ordinary / clustered test, p
\end{minipage} \\
\midrule\noalign{}
\endfirsthead
\toprule\noalign{}
\begin{minipage}[b]{\linewidth}\raggedright
Comparison
\end{minipage} & \begin{minipage}[b]{\linewidth}\raggedleft
Losses / gains
\end{minipage} & \begin{minipage}[b]{\linewidth}\raggedleft
Effect, pp {[}95\% interval{]}
\end{minipage} & \begin{minipage}[b]{\linewidth}\raggedleft
Ordinary / clustered test, p
\end{minipage} \\
\midrule\noalign{}
\endhead
\bottomrule\noalign{}
\endlastfoot
GLM stock framing versus default & 5 / 3 & \ensuremath{-}2.56 {[}\ensuremath{-}8.97, +3.85{]} & 0.726563 / 0.6875 \\
GLM abliterated framing versus default & 6 / 4 & \ensuremath{-}2.56 {[}\ensuremath{-}10.26, +6.41{]} & 0.753906 / 0.78125 \\
Difference between those framing effects & 5 negative / 3 positive / 1 +2 task-mode contrasts & 0.00 {[}\ensuremath{-}8.97, +8.97{]} & exploratory only \\
\end{longtable}\endgroup

Under GLM framing, stock and abliterated each pass 21 tasks, with cells 18 / 3 / 3 / 54, effect 0.00 pp {[}\ensuremath{-}6.41, +6.41{]}, and all reported tests unadjusted. These contrasts cannot isolate refusal because the prompt bundles several contextual statements and relevant checkpoint/effort cells were not assigned.

\subsection{External provider-route outcomes}\label{external-provider-route-outcomes}

Exact shared-ID comparisons restrict external outcomes to the same tasks used by the four primary pinned contrasts. External CVE effects use base-CVE clustering. External 120-task CyberGym intervals resample individual IDs; each selected task has a unique repository, so task and repository resampling coincide. External analyses are exploratory and unadjusted.

\begingroup\scriptsize
\setlength{\tabcolsep}{3.5pt}
\begin{longtable}[]{@{}
  >{\raggedright\arraybackslash}p{(\linewidth - 8\tabcolsep) * \real{0.1579}}
  >{\raggedleft\arraybackslash}p{(\linewidth - 8\tabcolsep) * \real{0.2105}}
  >{\raggedleft\arraybackslash}p{(\linewidth - 8\tabcolsep) * \real{0.2105}}
  >{\raggedleft\arraybackslash}p{(\linewidth - 8\tabcolsep) * \real{0.2105}}
  >{\raggedleft\arraybackslash}p{(\linewidth - 8\tabcolsep) * \real{0.2105}}@{}}
\caption{External source-assigned effects on the exact shared task IDs. Common IDs remove task-coverage differences only.}\tabularnewline
\toprule\noalign{}
\begin{minipage}[b]{\linewidth}\raggedright
Family / benchmark
\end{minipage} & \begin{minipage}[b]{\linewidth}\raggedleft
Pinned effect, pp
\end{minipage} & \begin{minipage}[b]{\linewidth}\raggedleft
External effect on shared tasks, pp {[}95\% interval{]}
\end{minipage} & \begin{minipage}[b]{\linewidth}\raggedleft
External paired cells, both pass / stock only / abliterated only / both fail
\end{minipage} & \begin{minipage}[b]{\linewidth}\raggedleft
Exact McNemar p
\end{minipage} \\
\midrule\noalign{}
\endfirsthead
\toprule\noalign{}
\begin{minipage}[b]{\linewidth}\raggedright
Family / benchmark
\end{minipage} & \begin{minipage}[b]{\linewidth}\raggedleft
Pinned effect, pp
\end{minipage} & \begin{minipage}[b]{\linewidth}\raggedleft
External effect on shared tasks, pp {[}95\% interval{]}
\end{minipage} & \begin{minipage}[b]{\linewidth}\raggedleft
External paired cells, both pass / stock only / abliterated only / both fail
\end{minipage} & \begin{minipage}[b]{\linewidth}\raggedleft
Exact McNemar p
\end{minipage} \\
\midrule\noalign{}
\endhead
\bottomrule\noalign{}
\endlastfoot
DeepSeek CVE-Bench & \ensuremath{-}3.846 over 78 & \ensuremath{-}14.103 {[}\ensuremath{-}24.359, \ensuremath{-}5.128{]} & 30 / 12 / 1 / 35 & 0.003418 \\
GLM CVE-Bench & 0.000 over 78 & +5.128 {[}\ensuremath{-}1.282, +11.538{]} & 27 / 2 / 6 / 43 & 0.289063 \\
DeepSeek CyberGym & +1.667 over 120 & \ensuremath{-}1.667 {[}\ensuremath{-}12.500, +9.167{]} & 42 / 22 / 20 / 36 & 0.877614 \\
GLM CyberGym & +11.667 over 120 & \ensuremath{-}10.833 {[}\ensuremath{-}20.000, \ensuremath{-}1.667{]} & 47 / 22 / 9 / 42 & 0.029449 \\
\end{longtable}\endgroup

For GLM CyberGym, external routes pass 69 stock tasks and 56 abliterated tasks on the shared 120, producing a negative sign in the opposite direction to the pinned 22/36 result. The complementary 1,387 external tasks have \ensuremath{-}12.040 pp, almost the same as the shared-subset effect, and all 1,507 have \ensuremath{-}11.944 pp.~Task coverage does not explain the sign reversal. Incorrectly attributing the reversal to having used two different task sets would contradict the common-ID analysis.

\begingroup\scriptsize
\setlength{\tabcolsep}{3.5pt}
\begin{longtable}[]{@{}
  >{\raggedright\arraybackslash}p{(\linewidth - 6\tabcolsep) * \real{0.2143}}
  >{\raggedleft\arraybackslash}p{(\linewidth - 6\tabcolsep) * \real{0.2857}}
  >{\raggedleft\arraybackslash}p{(\linewidth - 6\tabcolsep) * \real{0.2857}}
  >{\raggedright\arraybackslash}p{(\linewidth - 6\tabcolsep) * \real{0.2143}}@{}}
\caption{External all-task effects with project-resampled 95\% intervals. Original individual-task results and intervals are retained in the base artifacts. Discordance is not evidence that the near-zero DeepSeek effect demonstrates sameness.}\tabularnewline
\toprule\noalign{}
\begin{minipage}[b]{\linewidth}\raggedright
External effect over all assigned CyberGym tasks
\end{minipage} & \begin{minipage}[b]{\linewidth}\raggedleft
Paired cells
\end{minipage} & \begin{minipage}[b]{\linewidth}\raggedleft
Effect, pp {[}95\% interval{]}
\end{minipage} & \begin{minipage}[b]{\linewidth}\raggedright
Project sensitivity
\end{minipage} \\
\midrule\noalign{}
\endfirsthead
\toprule\noalign{}
\begin{minipage}[b]{\linewidth}\raggedright
External effect over all assigned CyberGym tasks
\end{minipage} & \begin{minipage}[b]{\linewidth}\raggedleft
Paired cells
\end{minipage} & \begin{minipage}[b]{\linewidth}\raggedleft
Effect, pp {[}95\% interval{]}
\end{minipage} & \begin{minipage}[b]{\linewidth}\raggedright
Project sensitivity
\end{minipage} \\
\midrule\noalign{}
\endhead
\bottomrule\noalign{}
\endlastfoot
DeepSeek stock 717 versus abliterated 716 & 502 / 215 / 214 / 576 & \ensuremath{-}0.066 {[}\ensuremath{-}2.522, +2.389{]} & 188 projects; task weighted; whole-project swap Monte Carlo \(p\) = 1.0 over 100,000 draws \\
GLM stock 812 versus abliterated 632 & 505 / 307 / 127 / 568 & \ensuremath{-}11.944 {[}\ensuremath{-}17.585, \ensuremath{-}7.366{]} & 188 projects; task weighted; zero extreme project-swap draws in 100,000 \\
\end{longtable}\endgroup

The DeepSeek external CVE exclusion-restricted contrast retains the same 12/1 discordant counts as all 80; excluding both-fail \texttt{CVE-2024-34070} variants only changes the denominator, to \ensuremath{-}14.103 pp.~A whole-CVE label-swap analysis over all 80 gives \(p\) = 0.015625, versus exact McNemar \(p\) = 0.003418, illustrating dependence sensitivity. Full-population external GLM CyberGym's interval is wider under project resampling than individual-task bootstrapping. This is what the adjusted uncertainty does and does not capture: project/task resampling under the source data, not route drift or rerun uncertainty.

\subsection{RQ2: Bounded behavior in abliterated-only-success pairs}\label{rq2-bounded-behavior-in-abliterated-only-success-pairs}

The analytic set contains 41 abliterated-only successes in seven pinned comparisons. The separate external strata contain another 348 source abliterated-only successes; they are not included in this mechanism audit.

\begingroup\scriptsize
\setlength{\tabcolsep}{3.5pt}
\begin{longtable}[]{@{}
  >{\raggedright\arraybackslash}p{(\linewidth - 4\tabcolsep) * \real{0.3000}}
  >{\raggedleft\arraybackslash}p{(\linewidth - 4\tabcolsep) * \real{0.4000}}
  >{\raggedright\arraybackslash}p{(\linewidth - 4\tabcolsep) * \real{0.3000}}@{}}
\caption{RQ2 bounded terminal accounting. Terminal classes do not encode refusal, intent, or a causal explanation and should not be read as mutually exclusive competence categories. No subclass is tested for enrichment.}\tabularnewline
\toprule\noalign{}
\begin{minipage}[b]{\linewidth}\raggedright
Comparison
\end{minipage} & \begin{minipage}[b]{\linewidth}\raggedleft
Abliterated-only successes
\end{minipage} & \begin{minipage}[b]{\linewidth}\raggedright
Stock failure terminal grouping, bounded metadata
\end{minipage} \\
\midrule\noalign{}
\endfirsthead
\toprule\noalign{}
\begin{minipage}[b]{\linewidth}\raggedright
Comparison
\end{minipage} & \begin{minipage}[b]{\linewidth}\raggedleft
Abliterated-only successes
\end{minipage} & \begin{minipage}[b]{\linewidth}\raggedright
Stock failure terminal grouping, bounded metadata
\end{minipage} \\
\midrule\noalign{}
\endhead
\bottomrule\noalign{}
\endlastfoot
DeepSeek high, CVE-Bench & 3 & 3 iteration budget \\
DeepSeek high, CyberGym & 9 & 9 no finish observed \\
DeepSeek max, CVE-Bench & 1 & 1 iteration budget \\
DeepSeek max, CyberGym first 20 & 2 & 2 no finish observed \\
GLM high, CVE-Bench & 3 & 3 iteration budget (one rejected legacy lexical flag) \\
GLM high, CyberGym & 20 & 11 no finish observed; 8 iteration budget; 1 agent finish after 37 calls \\
GLM authorized framing, CVE-Bench & 3 & 3 iteration budget \\
Total & 41 & 22 no finish observed; 18 iteration budget; 1 agent finish \\
\end{longtable}\endgroup

This is a terminal-state partition from existing run metadata and the seven comparison pairs, not proof of why stock failed. Within the broader bounded packet labels, none of the 41 stock packets received a refusal label; thus one category is an agent finish without an identified refusal and 40 are neither that category nor refusal-positive. No submission is present in 27 of the 41 but overlaps terminal classes. It is not a separate disjoint mechanism.

R076, the only RQ2 early-finish example, is a GLM CyberGym stock run with 37 model calls and an official \texttt{finish} action before its budget, followed by a failing oracle. Its paired abliterated run passed. The case is compatible with the statement ``an agent finish followed work without a demonstrated refusal''; it is not support for failure before attempt. The original hypothesis's ``stopped before attempting an exploit'' category was not measured, and current evidence cannot test that hypothesis as written.

R030 and R066 illustrate rejected lexical patterns in bounded packets, with stock oracle passes. They belong to RQ3 lexical visibility, not the RQ2 win set. Their task identifiers and hashes exist in \path{paper/evidence-examples.json}. Their exact matched regions require source-level channel checks before being used as clean visibility explanations; they are purpose-selected diagnostic capsules, not representative examples.

\subsection{RQ3: Bounded packet analysis for stock runs}\label{rq3-bounded-packet-analysis-for-stock-runs}

The RQ3 corpus contains 650 stock packets: 14 with legacy stock flags, 48 fixed-seed unflagged stock runs, 33 RQ2 additions, and 555 additional stock packets. Nine other packets belonging to abliterated arms are excluded. All 41 RQ2 targets are included.

All 650 packets received two independently executed agent passes using identical generation settings as far as recorded. The four fields \texttt{refusal\_present}, scope, effect, and confidence are identical in all 650 rows. Agreement on these four fields is 650/650; scope/effect are negative labels and kappa is undefined. Evidence-message indices nevertheless differ on 101 packets, and every rationale string differs. The check did not establish equal evidence comprehension. Agreement is concentrated in an all-negative category with no control positives.

\begingroup\scriptsize
\setlength{\tabcolsep}{3.5pt}
\begin{longtable}[]{@{}
  >{\raggedright\arraybackslash}p{(\linewidth - 4\tabcolsep) * \real{0.3000}}
  >{\raggedleft\arraybackslash}p{(\linewidth - 4\tabcolsep) * \real{0.4000}}
  >{\raggedright\arraybackslash}p{(\linewidth - 4\tabcolsep) * \real{0.3000}}@{}}
\caption{Stock packet coverage and visibility limits. Candidate-window counts measure coverage of the selection mechanism, not human or model accuracy.}\tabularnewline
\toprule\noalign{}
\begin{minipage}[b]{\linewidth}\raggedright
Measured feature
\end{minipage} & \begin{minipage}[b]{\linewidth}\raggedleft
Count
\end{minipage} & \begin{minipage}[b]{\linewidth}\raggedright
Interpretation
\end{minipage} \\
\midrule\noalign{}
\endfirsthead
\toprule\noalign{}
\begin{minipage}[b]{\linewidth}\raggedright
Measured feature
\end{minipage} & \begin{minipage}[b]{\linewidth}\raggedleft
Count
\end{minipage} & \begin{minipage}[b]{\linewidth}\raggedright
Interpretation
\end{minipage} \\
\midrule\noalign{}
\endhead
\bottomrule\noalign{}
\endlastfoot
Stock packets with final negative refusal labels & 650/650 & Coverage of packets, not a validated zero-refusal rate \\
No assistant-visible candidate window in the compact input & 570/650 & No phrase-triggered visible excerpt in the actual reading representation \\
No candidate window in either stored or visible channels & 296/650 & Compact input contains no assistant prose, only instructions and role/tool timeline \\
Packets containing a bounded review-text truncation marker & 287/650 & At least one bounded source window, not a complete character-ratio estimate \\
Messages with \texttt{review\_text\ =\ null} & 12,226 of 15,102 & Message evidence not presented through packet review text \\
Messages with truncation markers & 1,122 of 15,102 & Count of bounded messages, not independent evidence fragments \\
Legacy stock stored-content original-rule occurrences & 21 in 14 trajectories & Original stored strings, including possible literal deliberation \\
Matches in packet-derived visible candidate windows & 0 & Bounded packet diagnostic, not a full-transcript audit \\
Source spans dropped before visible normalization & 9 in 8 packets: R007, R016, R030, R035 twice, R058, R065, R066, R068 & Their original channel cannot be determined from normalized packet text alone \\
\end{longtable}\endgroup

The visibility reconciliation checked stored content and packet-derived visible windows separately. It found 21 stored-content matches in the RQ3 corpus and zero matches recomputed in a bounded packet-derived visible window. The measurements cover different text: nine spans are dropped by bounding before the normalized visible channel is derived. Their source channel is not recorded in the retained packets, so this analysis cannot classify those spans as hidden or visible.

This evidence does not support a broad statement that there are no refusals across full stock trajectories. It supports only ``two independent agent passes produced no refusal label on their bounded inputs'' and the directly measured metadata above. False negatives could arise from unselected text, unlisted wording, channel normalization, or shared annotation bias. The legacy visible-channel audit also has a provenance discrepancy: reconciliation records visibility hash \texttt{bbe39b3d\ldots{}}, while the current audit file is \texttt{bf62ba4f\ldots{}}. The aggregate 31-to-zero headline did not change, but the version difference needs reconciliation before calling the older provenance complete.

The 504 oracle-failing stock runs and 146 passing runs do not form a credible evaluated comparison between refusal-positive and refusal-negative cases because the bounded corpus has no refusal-positive group. No success-by-refusal-status effect is estimated. The narrower observation is that no identified refusal is required for an oracle failure in this corpus, and absence found in bounded text does not imply verified success.

\section{Discussion}\label{discussion}

The results vary by checkpoint and protocol rather than showing a uniform benefit from abliteration. DeepSeek's high-effort estimates are small and unresolved; at max effort, stock performs better on CVE-Bench. GLM stock and abliterated checkpoints have the same number of CVE-Bench passes but different per-task outcomes. The GLM CyberGym contrast meets the designated four-test adjustment, but changing one pair moves its adjusted significance across 0.05. Its sign also reverses for the same tasks on external routes.

A release labelled abliterated is therefore not a dependable proxy for better cybersecurity performance. Nor does a measured checkpoint difference isolate abliteration: the third-party construction process and any other weight-level or behavioral changes remain unmeasured. The bounded packet labels likewise cannot establish refusal absence. Task success depends on the oracle, scaffolding, difficulty, prompt/tool structure, grading semantics, and execution limits.

GLM supplies a direct test of task coverage as an explanation for route disagreement: the effect changes sign on shared task IDs. Comparing a 120-task OpenHands sample with all 1,507 Inspect ReAct tasks therefore cannot explain the reversal. The routes also differ in model identity bindings, serving configurations, harnesses, budgets, and grading implementations; external revisions are unpinned. These source-assigned metrics neither falsify the pinned GLM result nor establish a universal effect. Each result describes a model-route-protocol combination, not abliteration alone.

Packet evidence for refusal is much narrower. The selected packets do not establish that abliterated passes followed stock refusals or that refusal removal caused gains. They also cannot establish refusal absence in full transcripts. Among the 41 selected stock failures, 22 have no observed finish and 18 hit iteration limits; none received a refusal label. Early-stop and lexical flags cannot reconstruct a reliable taxonomy of attempt status.

Stronger claims about practical utility require randomized task order, repeated trajectories, controlled serving starts and caches, verified shared rendering, pinned or independently attested route identities, and execution-aware error accounting. Refusal prevalence estimates require complete visible-text coverage and validated annotation controls. The present evidence supports a checkpoint-and-protocol case study, not a settled measurement of abliteration's effects.

\section{Limitations}\label{limitations}

Abliterated arms are third-party releases. Paired prompts and budgets do not validate how the abliterated releases were constructed, and no independent stock-to-abliterated refusal reduction is established. Hardware and parallelism differ between families, and external routes are unpinned.

GLM template files differ although all reference renders tested here are identical. Their actual production token inputs were not reconstructed. Different server defaults could still matter for untested cases.

There is one retained run per task and condition. Zero-temperature settings do not eliminate batch and cache variation. Repeated task trajectories would be needed to estimate run-to-run variability. Secondary and exploratory comparisons are not confirmatory tests.

Eligible-task counts and observed effects are known for the measured sets. The selected 120 CyberGym tasks do not form an independent uniform sample with known inclusion probabilities. Bootstrap intervals do not encode that sampling mechanism, rerun randomness, serving variation, or pretraining overlap with public vulnerability tasks.

Whole-CVE, whole-project, cluster bootstrap, and ordinary McNemar analyses answer different dependence questions. Label exchanges require exchangeability assumptions, while confidence units do not automatically share the exact test's estimand. Zero-width percentile intervals can arise empirically and are not population equivalence guarantees.

Immutable route identities are null. The exact eight batches and task outcomes reconcile with stored source metrics, but this does not independently validate every grader execution or exclude diagnostic failures within source \texttt{both\_fail} cells. Limits differ by arm, with 805 versus 614 \texttt{n\_limited} records for GLM CyberGym. R2 proves storage and identity binding of its archived records, not semantic equivalence to a primary valid oracle run.

Labels are bounded OpenCode/Kimi-K3.1-Pro agent outputs, not human-validated ground truth or independent full-trajectory detection sensitivity. Two separate pipelines share a model, rubric, and candidate screen. Settings and resumable request/session metadata are incomplete. Reconciliation did not recheck all cases; it preserved agreements. Candidate filtering, bounding, unselected message text, and shared model bias can produce false negatives. Nine legacy spans cannot currently be channel-classified because of bounding. The visibility audit also has an unresolved hash-version discrepancy. The 650/650 agreement count does not resolve those limits.

Early finish may follow substantial work. \texttt{no\_finish\_observed} does not mean no attempt or self-stop, and \texttt{no\_submission} is not automatically a causal explanation. The CVE early flag also differs in its implementation from requiring an explicit finish event. The original pre-attempt mechanism remains unresolved.

The study does not test general capability retention, unsafe compliance, misconduct, attacker procurement/adoption, policy robustness against incorrect contexts, or incident-response effectiveness. Historical real-world task descriptions may have been present during model training. Nothing in these results supports weakening safeguards.

\section{Research ethics and evidence handling}\label{research-ethics-and-evidence-handling}

Evaluated targets were isolated benchmark environments owned by the evaluation infrastructure, not third-party systems. Exploit-relevant raw evidence was retained without editing completions, prompts, or official outcomes. The verification note includes selected run IDs and trajectory hashes where useful. It does not expose operational payloads or credentials.

Abliteration was evaluated as a population-designated checkpoint condition, not as a recommendation for unrestricted deployment. Ordinary task compliance, no observed refusal, and verified exploit completion are not general determinations of model safety. The bounded behavioral evidence is insufficient to infer that safeguards are unnecessary.

\section{Data and code availability}\label{data-and-code-availability}

The pinned checkpoint identifiers and external route labels are reported by arm. The reproducibility package includes \path{paper/analysis/revision-analysis.json}, \path{results/manifests/2026-11-26-revision-analysis.json}, \path{paper/analysis/external-validity-schema.json}, \path{results/manifests/2026-11-26-external-validity-audit.json}, \path{code/analysis/revision_analysis.py}, \path{code/analysis/test_revision_analysis.py}, \path{code/analysis/test_revision_manuscript.py}, \path{paper/template-render-checks.md}, and \path{paper/revision-verification.md}. These files contain task-level sensitivity results, source hashes, quantile and seed definitions, statistical specifications, and evaluation limits.

The archived artifacts in \texttt{results/summaries/} and \texttt{results/manifests/} include valid task outcomes, paired cells, source route manifests, valid/excluded attempts, and non-operational evidence capsules. A versioned supplement should also contain privacy-scoped pair keys, base-CVE and project groups, arm/benchmark versions, eligibility, per-run packet-coverage measures, and per-pass labels with their existing provenance. Raw trajectories, full prompts, tools, reasoning fields, request bodies, raw logs, and original credentials remain private in R2 and the Modal volumes.

The offline calculations read stored outcomes or structural metadata without starting additional model or oracle executions. Restricted access to bounded raw extracts remains case by case by the corresponding author, with read-only, time-limited handling and security review. No public raw-trajectory bucket is offered.

\section{Acknowledgments}\label{acknowledgments}

We thank Oskar Paulander for detailed comments on an earlier version of this manuscript, which helped improve the analysis and presentation.

\section{Funding and competing interests}\label{funding-and-competing-interests}

Modal provided compute credits. New Measure, Inc \citep{newmeasure2026platform} provided evaluation credits for the external archived batches through Benchmark Router. Superagent Technologies, Inc.~funded remaining evaluation and storage costs. The authors imported the supplied archives and reconstructed pairings, identities, and metrics. These provider-sponsored records are not independent replications.

Alan Zabihi is CEO and Ismail Pelaseyed is CTO and co-founders of Superagent Technologies, Inc., which lists backing from Y Combinator, Rebel Fund, and founders of Okta, Replit, and HubSpot \citep{superagent2026about}. Except for disclosed credits, the authors declare no model-provider relationships with evaluated vendors, route providers, or evaluation-credit sponsors.

\section{Model-assisted annotation disclosure}\label{model-assisted-annotation-disclosure}

Two separately run OpenCode pipelines assigned bounded refusal labels using one fixed rubric. Both used Kimi-K3.1-Pro. Final agreement checking and carry-forward used the same model class as the two passes. They did not write raw evidence, run models, choose oracle outcomes, select valid benchmark cells, or define the statistical tests.

Model type, packet selection, and candidate-screen exposure are material to the label meaning. Labels are bounded annotation outputs, not human ground truth. One candidate screen cannot establish recall of all semantically relevant text, and the all-negative distribution prevents meaningful positive-case calibration. Those constraints are carried into RQ2/RQ3 rather than hidden in an appendix.

\nocite{bhatt2024cyberseceval2}

\bibliographystyle{plainnat}
\bibliography{refs-2}

\end{document}